\documentstyle [12pt,a4,epsf] {article}
\newcommand{\e} { {\rm e} }

\newcommand{\DA} {D_{\rm A}}
\newcommand{\DB} {D_{\rm B}}
\newcommand{\eq} {0,{\rm eq}}
\newcommand{\seq} {0}

\newcommand{\ap} {\phi}
\newcommand{\bp} {\psi}
\newcommand{\water} {\eta}

\newcommand{\moa} {\mu_{1,{\rm A}}}
\newcommand{\mob} {\mu_{1,{\rm B}}}

\newcommand{\mza} {\mu_{0,{\rm A}}}
\newcommand{\mzb} {\mu_{0,{\rm B}}}

\newcommand{\mba} {\mu_{{\rm b},{\rm A}}}
\newcommand{\mbb} {\mu_{{\rm b},{\rm B}}}

\newcommand{\ala} {\alpha_{\rm A}}
\newcommand{\alb} {\alpha_{\rm B}}
\newcommand{\bta} {\beta_{\rm A}}
\newcommand{\btb} {\beta_{\rm B}}

\newcommand{\toa} {\tau_{1,{\rm A}}}
\newcommand{\tob} {\tau_{1,{\rm B}}}
\newcommand{\toab} { \bar{\tau}_{1,{\rm A}} }

\newcommand{\tza} {\tau_{0,{\rm A}}}
\newcommand{\tzb} {\tau_{0,{\rm B}}}
\newcommand{\tzab} { \bar{\tau}_{0,{\rm A}} }

\newcommand{\tgab} { \bar{\tau}_{\gamma,{\rm A}} }
\newcommand{\tgbb} { \bar{\tau}_{\gamma,{\rm B}} }

\begin{document}


\title {Kinetics of Surfactant Adsorption at Fluid-Fluid Interfaces:
             Surfactant Mixtures }
\author {Gil Ariel, Haim Diamant and David Andelman$^*$ \\
         \\
         School of Physics and Astronomy   \\
         Raymond and Beverly Sackler Faculty of Exact Sciences \\
         Tel Aviv University, Ramat Aviv, Tel Aviv 69978, Israel \\
         \\}

\date{March 1999}
\maketitle


\begin{abstract}
\setlength {\baselineskip} {16pt}

The adsorption at the interface between an aqueous solution of
several surface-active agents and another fluid (air or oil) phase
is addressed theoretically.
We derive the kinetic
equations from a variation of the interfacial free energy, solve them
numerically and provide an analytic solution for the simple case of a
linear adsorption isotherm.
Calculating asymptotic solutions analytically, 
we find the characteristic time
scales of the adsorption process and observe the behavior of the system
at various temporal stages.
In particular, we relate the kinetic behavior of the mixture to the
properties of its individual constituents and find good agreement
with experiments.
In the case of kinetically limited adsorption, 
the mixture kinetics is found to be considerably different from that 
of the single-surfactant solutions because of
strong coupling between the species.

\end{abstract}

\vspace{5cm}
To be published in {\it Langmuir}.


\pagebreak
\section{Introduction}
\label{introduction}
\setcounter{equation}{0}
\setlength {\baselineskip} {16pt}

The kinetics of surfactant adsorption plays an important
role in various
interfacial phenomena and has been an active field of research,
both experimentally and theoretically,
since the 40s \cite{review}.
Recently, a new theoretical approach based on a
free-energy formalism was presented  and applied to
nonionic as well as ionic surfactant adsorption \cite{Hpaper}.
The main advantage of this approach is that all the
equations are derived from a single functional.
This feature facilitates generalizations of the model to more
complicated situations.

Surfactant mixtures are used in numerous industrial
applications, and are also encountered in many systems
because of the presence of surface-active impurities.
Experiments portray a
large variety of phenomena 
specific to the kinetics of
mixed systems \cite{Fainerman1}--\cite{mercury}. 
For instance,
more complex dynamic surface tension is observed due to
competition between the different species. 

The equilibrium behavior of mixed-surfactant solutions
was studied in detail in previous works 
\cite{Lucassen}--\cite{Blank2}.
One of the important results, both theoretically and from 
the application
point of view, is the ability to relate the mixed-surfactant 
behavior
to that of the better understood, single-surfactant one.
In the current work we focus on kinetic aspects,
deriving equilibrium results merely for completeness and comparison
with previous models.
One of our aims is to predict the
mixture kinetics from the behavior of the single surfactants.
A particularly interesting question is whether 
mixing several species would lead in certain cases 
to a significant difference
in the kinetics as compared to the single-surfactant systems.

In this paper we extend our previously
introduced model to describe
the competition between two nonionic adsorbing species.
In sec.~\ref{motion} we present the model and derive the
equilibrium relations and kinetic equations.
The complete set of kinetic equations can be solved only numerically, 
as is done
in sec.~\ref{solution}.
We then discuss, in sec.~\ref{asymptotic}, the asymptotic time dependence
of diffusion-limited and kinetically limited adsorption and the corresponding
characteristic time scales.
We focus on the relation between the adsorption behavior of the mixture
and the properties of its individual constituents.
Good agreement is found between
the experimental results and our predictions.
In addition, we give an
analytic solution for the kinetics in 
the simple case of a linear adsorption
isotherm.

In order to make the formulation as transparent as
possible, we have tried to minimize the number of symbols
and used dimensionless forms whenever possible.
To facilitate translation into experimentally useful,
dimensional quantities, a nomenclature table is provided in
Appendix B.


\section{The Model}
\label{motion}
\setcounter{equation}{0}

Our model is based on the free-energy formalism presented in detail
in previous papers \cite{Hpaper}.
We consider a semi-infinite aqueous
solution of nonionic surfactants having, at $x=0$, a flat, sharp 
interface
with a non-polar fluid phase (such as air or oil). The solution is 
in contact
with a bulk reservoir, at $ x \rightarrow \infty$, containing
two types of surfactant molecules, denoted by A and B.
It is considered dilute, with the volume fraction of the two 
constituents
well below their critical micelle concentration (cmc) values 
\cite{micelle}. At the
interface, however,  the surfactant volume fraction is usually much 
larger.
Hereafter we use $\ap$ to denote volume fraction of surfactant A,
$\bp$ for volume fraction of surfactant B,
$\mu$ for chemical potential
and $D$ for diffusion constant. The subscripts ${\rm A}$ and ${\rm B}$
distinguish between the two different surfactants.
The subscripts $0,1,{\rm b}$ are used to distinguish between different positions
in the solution, corresponding, respectively, to the interface,
sub-interface layer and bulk reservoir.

The excess interfacial free energy
is equal to the reduction in
surface tension,
\begin{equation}
   \Delta\gamma[\ap, \bp ] = \int_{0}^{\infty} \left\{
 \Delta f [\ap(x) ] + \Delta f [ \bp
(x) ] \; \right\} {\rm d}x +  f_{0} (\ap_{0} , \bp_{0} )
\label{a13}
\end{equation}
where $\Delta f$ is the bulk
contribution per unit volume of each species, and
$ f_{0} $ the interfacial contribution per unit area. 
In dilute nonionic surfactant systems, the dominant
contribution to the surface tension is usually the
interfacial one, $f_0$.
Since the
surfactant solution is considered dilute,
the bulk free energy is taken to be the sum of
$\Delta f(\ap) $ and
$\Delta f(\bp) $.
Each of these contributions contains the entropy of mixing in the
ideal, dilute limit and a contact with the bulk reservoir.
For species A we write
\begin{equation}
\Delta f(\ap ) = \frac{1}{a^3} \left[ \ap \ln
\ap - \ap  -
( \ap_{\rm b} \ln \ap_{\rm b} - \ap_{\rm b})
 - \mba
 ( \ap - \ap_{\rm b} ) \right]
\label{a15}
\end{equation}
and similarly for species B.
As this is a lattice model, it is conveniently formulated
using volume fractions as the degrees of freedom.
The translation to mole fractions or concentrations requires
specifying molecular sizes for the various species.
Here, the surfactant molecular size, $a$, is assumed to have the
same value for both species, on account of simplicity \cite{DifA}.
All the energies, chemical potentials and interaction parameters
are expressed in units of $ k_{\rm B} T $ where $ k_{\rm B} $ is
the Boltzmann constant and $ T $ the temperature.
At the interface, the surfactant volume fractions are usually much 
larger and we have to consider 
the full, non-ideal entropy of mixing and additional 
interaction terms,
\begin{eqnarray}f_{0} (\ap_{0} , \bp_{0} )   =
  \frac{1}{a^2} \left[ \ap_{0}
 \ln \ap_{0} + \bp_{0} \ln \bp_{0}
   + \water_{0} \ln \water_{0}
 - (\ala+ \moa)\ap_{0} - \right.
(\alb + \mob)\bp_{0} \nonumber \\
\left. - \frac{ \bta }{ 2 } { \ap_{0} }^{2} -
\frac{ \btb }{ 2 } { \bp_{0} }^{2} -
\varepsilon \ap_{0} \bp_{0} \right]
\label{a14}
\end{eqnarray}
where $\alpha$ is the energy gain of adsorption onto
the interface, $\beta$
the interaction energy between molecules of the same kind, $\varepsilon$
the interaction
between different surfactants \cite{GammaExp}, and
$ \water_{0} \equiv 1 - \ap_{0} - \bp_{0} \; $
is the surface coverage of the solvent (water).
The interface is in contact with the sub-surface layer,
having a chemical potential $\mu_1 \equiv \mu(x\rightarrow 0)$.
Out of
equilibrium $\mu_1$ may differ from the bulk chemical
potential, $\mu_{\rm b}$. Although both $\alpha$ and $\mu_1$
are linearly coupled with the surface coverage, their physical origin
is quite different. 
The former is constant in time, characterizing the surface
activity of the specific surfactant and mainly dependent on the 
molecular structure (number of hydrocarbon groups in the surfactant
tail).
The latter is a time dependent
function participating in the surface kinetics.
Variation of $\Delta\gamma$ with respect to $\phi$ and $\psi$ yields the excess chemical
 potential at distance $x$ from the interface,

\begin{eqnarray}
   \Delta\mu_{\rm A} (x) = \ln \ap (x) - \mba \nonumber \\
   \Delta\mu_{\rm B} (x) = \ln \bp (x) - \mbb
\label{a16}
\end{eqnarray}
and at the interface itself,
\begin{eqnarray}
\Delta\mza =  \ln \frac{ \ap_{0} }{ \water_{0} }
    - \ala - \bta \ap_{0}  - \varepsilon \bp_{0} - \moa \nonumber \\
\Delta\mzb =  \ln \frac{ \bp_{0} }{ \water_{0} }
    - \alb - \btb \bp_{0}  - \varepsilon \ap_{0} - \mob
\label{a17}
\end{eqnarray}

\subsection{Equilibrium} 

At equilibrium the chemical
potentials are equal to their bulk values throughout the solution,
leading to two uniform profiles, $ \ap (x>0) \equiv \ap_{\rm b}$  and
$ \bp (x>0) \equiv \bp_{\rm b}$.
At the interface we obtain
a Frumkin adsorption 
isotherm \cite{adamson}, generalized for the A/B mixture case:

\begin{eqnarray}
\ap_{0} = \frac{ \ap_{\rm b} ( 1 - \bp_{0} ) }{ \ap_{\rm b}
   + \e^{  -( \ala + \bta \ap_{0} + \varepsilon \bp_{0} )  } } \nonumber \\
\nonumber \\
\bp_{0} = \frac{ \bp_{\rm b} ( 1 - \ap_{0} ) }{ \bp_{\rm b}
   + \e^{  -( \alb + \btb \bp_{0} + \varepsilon \ap_{0} )  } }
\label{a18}
\end{eqnarray}
The adsorption of species A depends on the species B
because of the
 entropy of mixing (steric effect) and surfactant-surfactant interactions.
The corresponding generalized Langmuir
isotherm is obtained in the limit of no interactions,
$\bta=\btb=
\varepsilon=0$.
Finally, the equilibrium equation of state, $ \Delta \gamma = \Delta
\gamma ( \ap,\bp) $, takes the form
\begin{equation}
   \Delta\gamma = \frac{1}{ a^{2} } \left(   \ln \water_{0} +
\frac{ \bta }{ 2 } {\ap_{0}}^{2} +
 \frac{ \btb }{ 2 } {\bp_{0}}^{2} + \varepsilon
\ap_{0} \bp_{0}  \right)
\label{a19}
\end{equation}
which is equivalent to Lucassen-Reynders' result 
\cite{Lucassen}, when
differences in molecular sizes are neglected.

\subsection{Out of Equilibrium}

We apply
the procedure presented in Ref.~\cite{Hpaper}, 
where the kinetic equations are derived from
the variation of the free energy. The procedure is generalized
for the present case of a two nonionic surfactant mixture.
Since the bulk solution is dilute,
two independent diffusion equations for the two
surfactants are obtained, 
leading to two Ward-Tordai equations \cite{WT},
similar to previous models \cite{FainermanMiller}:
\begin{eqnarray}
   \ap_{0} (t) = \frac{1}{ a } \sqrt{\frac{\DA}{\pi}}
\left[ 2\ap_{\rm b} \sqrt{t} - \int_{0}^{t} \frac{ \ap_{1}
(\tau) }{ \sqrt{t-\tau} }{\rm d}\tau  \right] + 2\ap_{\rm b} - \ap_{1} \nonumber \\
\nonumber \\
   \bp_{0} (t) = \frac{1}{ a } \sqrt{\frac{\DB}{\pi}}
\left[ 2\bp_{\rm b} \sqrt{t} - \int_{0}^{t} \frac{ \bp_{1}
(\tau) }{ \sqrt{t-\tau} }{\rm d}\tau  \right] + 2\bp_{\rm b} - \bp_{1}
\label{a11}
\end{eqnarray}
At the interface, however, the two species are correlated
and the procedure yields two coupled kinetic equations:
\begin{eqnarray}
   \frac{ \partial \ap_{0} }{\partial t} =
\ap_{1} \frac{ \DA }{ a^{2} } \left[ \ln \left( \frac{
\ap_{1} }{ \ap_{0} } \water_{0}  \right) + \ala  +
\bta \ap_{0} + \varepsilon \bp_{0}  \right] \nonumber \\
\nonumber \\
   \frac{ \partial \bp_{0} }{\partial t} =
\bp_{1} \frac{ \DB }{ a^{2} } \left[ \ln \left( \frac{
\bp_{1} }{ \bp_{0} } \water_{0}  \right) + \alb  +
\btb \bp_{0} + \varepsilon \ap_{0}  \right]
\label{a12}
\end{eqnarray}
As can be seen from the equation above,
the coupling between the kinetics of the two species
arises from an interaction term
as well as an entropic one (via the $\water_0$ term).
It should be mentioned that this form of interfacial
kinetic equations is particular to our free-energy approach.
Close enough to equilibrium, however, it coincides with the
usual adsorption-desorption form, as discussed in 
ref.~\cite{Hpaper}.
The system of four equations, (\ref{a11}) and (\ref{a12}), with the
appropriate initial conditions, completely determines the
mixture kinetics and equilibrium state. 
Several limits, such as diffusion-limited and kinetically
limited adsorption, can be treated analytically, as presented
in sec.~4.
The full solution
of the mixed-kinetics case is obtained numerically.


\section{Numerical Solution of the Full Equations}
\label{solution}
\setcounter{equation}{0}

Several numerical
schemes have been proposed for solving the Ward-Tordai equation
with various boundary conditions \cite{numeric}--\cite{Lin}.
We generalized the recursive scheme suggested by
Miller {\em et al.} \cite{numeric} to a surfactant mixture
having time dependent boundary conditions.
The complete set of integro-differential equations
were solved, as is explained in the Appendix.

The time dependence of $\ap_0$, $\bp_0$ and their sum $\ap_0+\bp_0$
can be seen in Fig.~1.
The mixture parameters are specifically chosen
to show the interesting case of competition
between the two species. While surfactant B diffuses more
rapidly and is more abundant at the interface during the initial
stages of the adsorption process, surfactant A has a higher
surface affinity and dominates the later stages of
adsorption. We note that, because of this competition,
surfactant A not only takes over the adsorption at the
later time stages, but it also forces surfactant B to desorb from
the interface. In Fig.~2 the dynamic surface tension
is shown for the same time scales and adsorption parameters
as of Fig.~1. The competition between the A and B surfactants
results in a more complex decrease of the surface tension at intermediate
times.


\section{ Limiting Time Behavior}
\label{asymptotic}
\setcounter{equation}{0}

The asymptotic time behavior close to equilibrium of the kinetic equations
provides a means for comparing theoretical predictions
to experimental results \cite{Fainerman1}. Two
limiting cases of adsorption can be distinguished
and treated separately for long and short times:
{\em Diffusion-Limited Adsorption} (DLA) occurs when the kinetics at
the interface is much faster than the diffusion from the bulk to the
sub-surface layer.
Equation~(\ref{a12}) then equals zero \cite{Hpaper}, since the interface
is taken to be at equilibrium with the sub-surface layer.
The profiles in the bulk, $\ap(x>0)$ and $\bp(x>0)$, depend on the diffusion
and evolve in time.
On the other hand,
{\em Kinetically Limited Adsorption} (KLA) takes place in the opposite
limit, for which
$\ap_{1} = \ap_{\rm b} $ and $\bp_{1}=\bp_{\rm b} $ throughout the process,
implying that the bulk is constantly at equilibrium with the reservoir but not
with the interface.

\vspace{0.5cm}
\subsection{ DLA process at long times }

Diffusion-limited processes such as the DLA considered here
have a characteristic asymptotic $t^{-1/2}$ dependence as $t$ goes
to infinity. This is demonstrated in Fig.~3.
The asymptotic time dependence of all variables can be
written in a generic form as

\begin{eqnarray}
\phi_0(t) & \simeq & \phi_{\eq} \left( 1 - \sqrt{ { \tau_{0} }/{ t } } \right)
\nonumber\\
\phi_1(t) & \simeq & \phi_{\rm b} \left( 1 - \sqrt{ { \tau_{1} }/{ t } } \right)
\nonumber\\
\Delta \gamma(t) & \simeq & \Delta \gamma_{\rm eq}
\left( 1 - \sqrt{{ \tau_{\gamma} } /{ t } }
\right)
\label{AsyForm}
\end{eqnarray}
where the subscript `eq' stands for equilibrium values.
Note that the various time constants may generally depend
on the surface coverages of both species.
Taking the limit $ t \rightarrow \infty $ of the Ward-Tordai
relation (\ref{a11}) we get

\begin{equation}
\toa =
 \frac{ a^2 }{ \pi \DA }
\left( \frac{ \ap_{\eq} }{ \ap_{\rm b} } \right)^2
\label{tau1}
\end{equation}
The presence of surfactant B changes the equilibrium value of
$\ap_0$ [see eq.~(\ref{a18})]. This is the {\it only} reason why the relaxation
time $\toa$ depends on surfactant B.
For a DLA process, the surface and sub-surface layers are
considered at equilibrium with each other, and
the coverage instantaneously follows any change in
the sub-surface concentration according to
\begin{eqnarray}
\ap_{0} = \frac{ \ap_{1} ( 1 - \bp_{0} ) }{ \ap_{1}
+ \e^{ -( \ala + \bta \ap_{0} + \varepsilon \bp_{0} ) } } \nonumber \\
\nonumber \\
\bp_{0} = \frac{ \bp_{1} ( 1 - \ap_{0} ) }{ \bp_{1}
+ \e^{ -( \alb + \btb \bp_{0} + \varepsilon \ap_{0} ) } }
\label{interface}
\end{eqnarray}
We now omit the subscript `eq' for brevity.
Substituting (\ref{AsyForm})
into (\ref{interface}) and comparing powers of $t$, we get
two linear equations for $ \tza $ and
$ \tzb $

\begin{eqnarray}
 \water_{\seq} \sqrt{ \toa }
&=&\left( 1 - \bp_{\seq} -
\bta \ap_{\seq} \water_{\seq}
  \right) \sqrt{ \tza } ~~ +~~
 \bp_{\seq} \left( 1 - \water_{\seq}  \varepsilon  \right)
\sqrt{ \tzb } \nonumber \\
 \water_{\seq} \sqrt{ \tob } & = &
\left( 1 - \ap_{\seq} - \btb \bp_{\seq} \water_{\seq}
 \right) \sqrt{ \tzb } ~~+~~
 \ap_{\seq} \left( 1 - \water_{\seq}  \varepsilon  \right)
 \sqrt{ \tza }
\label{linear}
\end{eqnarray}
Let us consider a single-surfactant system (say the A) as
a reference state. This is simply achieved by setting
$\bp_{\eq} $ and $ \tzb $ to zero,  and denoting $\ap_{\seq} \rightarrow
\bar{\ap}_{\seq} $, $ \water_{\seq} \rightarrow \bar{\water}_{\seq}
= 1 - \bar{\ap}_{\seq} $. Consequently,
\begin{equation}
\sqrt{ \tzab } = \frac{ 1 - \bar{\ap_{\seq}} }{ 1 - \bta
\left( 1-\bar{\ap_{\seq}} \right) \bar{\ap_{\seq}}
  } \sqrt{ \toab }
\label{single}
\end{equation}
where the ``bars'' throughout the paper denote values obtained
for this single-component system.

In a DLA process, the equation of
state (\ref{a19}) holds also
out of equilibrium \cite{Hpaper}, since the equilibration
of the interface with the sub-surface layer is very fast
for both species.
Substituting (\ref{AsyForm}) into (\ref{a19}),
the asymptotic behavior of the
surface tension is obtained:
\begin{equation}
-a^2\Delta\gamma\sqrt{ \tau_{\gamma} } =
\left( \frac{ \ap_{\seq} }{ \water_{\seq} } -
 \bta \ap_{\seq}^{2} - \varepsilon
\ap_{\seq}
\bp_{\seq} \right) \sqrt{ \tza }
+ \left( \frac{ \bp_{\seq} }{ \water_{\seq} } -
 \btb \bp_{\seq}^{2} - \varepsilon
\ap_{\seq}
\bp_{\seq} \right) \sqrt{ \tzb }
\label{gamma}
\end{equation}

Without surface interactions ($ \bta = \btb
= \varepsilon = 0 $) only steric effects and surface activity
terms are taken into account. 
The Frumkin-like isotherm
(\ref{interface}) is simplified in this approximation 
to a Langmuir-like isotherm, and
eq.~({\ref{linear}) takes the form
\begin{eqnarray}
\sqrt{ \tza } =  \left( 1 - \ap_{\seq} \right)
\sqrt{ \toa } -  \bp_{\seq} \sqrt{ \tob } \nonumber \\
\sqrt{ \tzb } =  \left( 1 - \bp_{\seq} \right)
\sqrt{ \tob } -  \ap_{\seq} \sqrt{ \toa }
\label{tau0}
\end{eqnarray}
Using eqs.~(\ref{single})--(\ref{tau0}) we obtain a simple
expression, relating $\tau_{\gamma}$
of the mixture with those of each species separately,
$\tgab$ and $\tgbb$,
\begin{equation}
\Delta\gamma\sqrt{ \tau_{\gamma} } = 
\Delta\bar{\gamma}_{\rm A} (
\ap_{\seq} / \bar{\ap_{\seq}} )^{2}
\sqrt{ \tgab } + 
\Delta\bar{\gamma}_{\rm B} (
 \bp_{\seq} / \bar{\bp_{\seq}})^{2}
\sqrt{ \tgbb }
\label{mix}
\end{equation}
where $\Delta\bar{\gamma}_{\rm A}$ and $\Delta\bar{\gamma}_{\rm B}$
are the equilibrium reduction in interfacial tension of the single-surfactant
solutions.
This correspondence relates the time scale of the surface tension relaxation
in the mixture with the time scales corresponding to the individual species.
As was
pointed out previously, most common nonionic surfactants usually undergo
DLA. Thus, the above result provides a convenient way of predicting the
behavior of  multi-component
surfactant mixtures based on single-surfactant data.
In Table 1, we compare the predicted
$\sqrt{\tau_\gamma}$ of eq.~(\ref{mix})
with experimental results obtained by Fainerman
and Miller \cite{FainermanMiller} for a sequence of Triton~X
mixtures.  Based on single-surfactant values and
equilibrium isotherms for the mixture,
the two terms of eq.~(\ref{mix}) are calculated separately.
The agreement between theory and experiment
is quite good, although
experiments were limited to cases having one species dominating
the adsorption.  The last entry in the table corresponds
to a mixture of Triton
X-405 and Triton X-165. Here the predicted $\tau_\gamma$
deviates from the experimental one by 33\%.
Equilibrium measurements on this mixture reveal an increase in
the X-165 coverage in the presence of the X-405 \cite{FainermanMiller}, 
implying strong
interfacial interactions between the species.  
The deviation in the predicted kinetics in Table 1 
probably arise from those interactions,
which are not
taken into account by eq.~(\ref{mix}). 
It is possible to treat also 
the general case, including interactions between surfactants,
by using the full equations (\ref{linear})-(\ref{gamma}) 
instead of the simplified
one (\ref{mix}). Such a procedure, however, involves 
 three additional fitting parameters ($\beta_{\rm A}$, $\beta_{\rm B}$
and $\varepsilon$).
Nevertheless, as demonstraed in Table 1, the simple prediction 
(\ref{mix})
may be applicable to various experimental systems.

\vspace{0.5cm}
\subsection{ KLA process at long times }

Although most nonionic surfactants undergo a DLA process,
the adsorption of
some nonionic surfactants as well as ionic ones
is found to be kinetically limited.
For a KLA process, the bulk of the mixture
is assumed to be at equilibrium,
$\ap (x) = \ap_{\rm b}$
and $\bp (x) = \bp_{\rm b}$.
The equations governing the kinetics are now the interfacial ones, 
(\ref{a12}).

Asymptotic solutions at $ t \rightarrow \infty $ of first-order 
equations
such as (\ref{a12}) have an exponential form characteristic of KLA.
Linearizing eqs.~(\ref{a12}) about the equilibrium state, 
$\ap_{\eq}$ and $\bp_{\eq}$,
two time scales denoted $\tau_{+}$ and $\tau_{-}$ emerge 
($ \tau_{-} > \tau_{+}$).
These collective time scales correspond to the kinetics of a certain
combination of surfactant coverages,
\begin{eqnarray}
 C_1 \Delta \ap_0 + C_2 \Delta \bp_0 & \sim & \e^{ - t/\tau_{-} }
\nonumber \\
 C_3 \Delta \ap_0 +  C_4 \Delta \bp_0 & \sim & \e^{ - t/\tau_{+} }
\label{longKLAmix}
\end{eqnarray}
where $\Delta \ap_0 \equiv \ap_0-\ap_{\eq}$, 
$\Delta \bp_0 \equiv \bp_0-\bp_{\eq}$,  and 
$C_1 \ldots C_4$ are constants.
Since  $ \tau_{-} > \tau_{+} $, it is $ \tau_{-}$ which  limits the kinetics of the system.

In the simple case of no surface interactions
between A and B
( $ \bta = \btb = \varepsilon = 0$, yet keeping steric and surface
activity effects), the expressions
for $ \tau_{\pm} $ are
\begin{equation}
\frac{ 2 }{ \tau_{\pm} } = 
\frac{ 1 - \bp_{\seq} }{ \tau_{\rm A} } +
\frac{ 1 - \ap_{\seq} }{ \tau_{\rm B} }
\pm  \sqrt{ \left( \frac{ 1 - \bp_{\seq} }{ \tau_{\rm A} } +
\frac{ 1 - \ap_{\seq} }{ \tau_{\rm B} } \right)^2
-4\frac {  \water_0 }
{ \tau_{\rm A} \tau_{\rm B} }}
\label{TauK}
\end{equation}
where $ \tau_{\rm A} $ and $ \tau_{\rm B} $ are the KLA time scales of
the single surfactant case, defined as
\begin{eqnarray}
 \tau_{\rm A} = \frac{ a^2 }{ \DA } \left(
  \frac{ \ap_0 }{ \ap_{\rm b} } \right)^2
\e^ { - \ala } \nonumber \\
\tau_{\rm B} = \frac{ a^2 }{ \DB } \left(
  \frac{ \bp_0 }{ \bp_{\rm b} } \right)^2
\e^ { - \alb }
\label{TauA}
\end{eqnarray}
yet $\ap_{0}$ and $\bp_{0}$
are the equilibrium values for the {\it mixture}.
The behavior of the mixed system combines the single-surfactant
kinetics in a complicated manner.
We can gain some insight on this coupling
by considering some simple cases.

For the case where the interfacial kinetics of surfactant A is much
slower than that of  B,
$\tau_{\rm A} \gg \tau_{\rm B}, $ eqs.~(\ref{longKLAmix}) and
(\ref{TauK}) are simplified to
\begin{eqnarray}
( 1 - \ap_{\eq}) \Delta \ap_0 -  \bp_{\eq}  \Delta \bp_0
   \sim \e^{ -t/\tau_{-} } & ; & \tau_{-} =\tau_{\rm A} ( 1- \ap_0 )  / \water_0
   \nonumber \\
\Delta \bp_0  \sim \e^{ -t/\tau_{+} } & ; & \tau_{+} = \tau_{\rm B}/( 1 - \ap_0 )
\label{Ta_ll_Tb}
\end{eqnarray}
In the other limiting case, where the two species have similar time scales,
$\tau_{\rm A} \simeq \tau_{\rm B} $, we get
\begin{eqnarray}
\Delta \ap_0  -  \Delta \bp_0 
   \sim \e^{ -t/\tau_{-} } & ; & \tau_{-} =  \tau_{\rm A} /  \water_0 
   \nonumber \\
\ap_{\eq}  \Delta \ap_0 + \bp_{\eq} \Delta \bp_0 
   \sim \e^{ -t/\tau_{+} } & ; & \tau_{+} = \tau_{\rm A}
\label{Ta_eq_Tb}
\end{eqnarray}
The factor $1/\water_0$ in $\tau_{-}$ is quite interesting.
 Since the
equilibrium surface coverage of the solvent, $\water_0$, is
usually very small in surfactant systems, this factor implies that
the coupling in a surfactant mixture undergoing KLA may lead to a
significant reduction in adsorption rate. In this regime 
the mixture behavior may differ considerably from that of its 
individual
constituents. Because of the relatively  large factor of 
$1/\water_0$,  
the KLA time scale may exceed the DLA one and
 the adsorption would become kinetically limited.

\vspace{0.5cm}
\subsection{ Short time behavior }

{\it DLA regime.}~~~
Here we assume that the surface and sub-surface layers are already at
equilibrium and examine the system at time scales short
compared with the diffusion
mechanism.
Note that the DLA behavior cannot strictly start at $t=0$, since 
at that instance the
interface and sub-surface layers are not at equilibrium with each other.
Assuming a DLA time dependence of the form
\( \;
\ap_{0}(t) \simeq \mbox{const.} + \sqrt{  t /
\tau_{\rm A } }
 \),
the $\mbox{const.}$ is found to be roughly equal
to $2\phi_{\rm b}$.
In other words, only once
the surface coverage has exceeded a value of $2\phi_{\rm b}$,
indicating an almost complete depletion of the
sub-surface layer \cite{WT},
can one assume a process limited by diffusion
(the same argument applies to the second surfactant).
Figure 4 shows the surface coverage and
surface tension at this DLA stage 
plotted as function of $t^{1/2}$.
The linear dependence of the surface
coverages is evident, whereas the dynamic surface tension, 
having contributions from both species, exhibits 
a small deviation from the $t^{1/2}$ behavior.  
Since surfactant A has been taken to be much
slower than surfactant B, the $t^{1/2}$ behavior 
of the two surfactants overlaps only for a 
very short period of time.

\noindent
{\it Beyond the DLA regime.}~~~
At the very early time stages of the process, most of the molecules in the
sub-surface layer migrate rapidly to the interface.
 Only when the sub-surface
 layer becomes nearly depleted, do molecules from the bulk
start migrating towards the interface by a diffusive
mechanism.
Before diffusion sets in, 
the DLA assumption is invalid and the interfacial
kinetics must be considered explicitly.
To address these very early time stages we assume
that the bulk solution is still at its initial equilibrium state,
unperturbed by the presence of the interface.
The leading time behavior of the surface coverage is found
from (\ref{a12}) to be linear, as can be seen in Fig.~5,
\begin{eqnarray}
\ap_0 (t \rightarrow 0) \simeq \ap_{\rm b} + \frac{
\DA }{ a^2 }
  \ap_{\rm b} \ala t 
\nonumber \\ \nonumber \\
\bp_0 (t \rightarrow 0) \simeq \bp_{\rm b} + \frac{
\DB }{ a^2 }
  \bp_{\rm b} \alb t 
\label{AssympAt0}
\end{eqnarray}
Any correlation between the two species vanishes
during this initial stage since the surfactants
at the interface are
dilute and non-interacting.
We would like to emphasize that those very early time scales 
($10^{-11}$--$10^{-7}$ s)
are of no experimental interest. We study them as an analytic limit of the general
kinetic equations.

\vspace{0.5cm}
\subsection{ The linear isotherm limit: weakly adsorbing surfactants }

We present now a very simplified and restricted case which,
nevertheless, is of interest since it has an analytic
solution. The assumptions
for which this case can be treated are:
(i) one of the surfactants (say A) has a low surface
activity and undergoes a slow DLA process;
(ii) the second surfactant (B)
is assumed to be at equilibrium; and (iii) the two species do not
interact.
Since the surface coverage
of surfactant A is always very low,
its isotherm can be taken as linear in $\phi_1$ (Henry's law):
\begin{equation}
\ap_0 = \e^{ \ala } \ap_1
\label{Henry}
\end{equation}
as can be seen from taking the low-coverage limit of eq.~(\ref{a18}).
With the above assumptions, the time-dependent
surface coverages of the two species can be explicitly calculated
\cite{Franses}--\cite{Sutherland},
\begin{eqnarray}
\ap_0(t) & = & 
[ 1 - \tanh(\ala/2) \e^{ t /\tau} {\rm erfc}(  \sqrt{t/\tau} )]
 \ap_{\rm b} \e^{ \ala }
\nonumber \\ \nonumber \\
\bp_0(t) & = & \frac{ \bp_{\rm b} \left[ 1 - \ap_0(t) \right] }{ \bp_{\rm b} +
\e^{-\alb } }
\label{erfc}
\end{eqnarray}
where $ \tau \equiv (a^2/ \DA)( 1+ \e^{ \ala} )^2 $.
Note that in the limit $t\rightarrow 0$, the DLA process
starts at
$\phi_0(t=0) = { 2 \phi_{\rm b} }/( \e^{ - \ala } + 1 ) $
which is somewhat smaller than $2 \phi_{\rm b} $.
This result is consistent
with an early-stage dynamics, where the sub-surface layer is
entirely depleted, as was discussed in Sec. 4.3. Furthermore,
the linear relation (\ref{Henry}) is always valid for short enough times,
when the surface coverage is low
(with no further limiting assumptions),
as demonstrated in Fig.~6.


\section{ Conclusions }
\label{conclude}
\setcounter{equation}{0}

We have examined the behavior of
nonionic surfactant mixtures by
solving
numerically the governing kinetic
equations, as well as deriving analytical limiting cases.
A comprehensive description
of the different adsorption stages is obtained.
For mixtures of common, nonionic surfactants,
the adsorption process can be divided into four temporal stages.
At the very early times of the adsorption, the process begins with
a short stage, where the surface coverage and surface tension
change linearly with time because of interfacial
kinetics (Fig.~5).
As discussed is sec. 4.3, this stage is practically too short
to be observed experimentally (usually less than microseconds).
During this early stage, the sub-surface layer becomes nearly empty,
which in turn drives a second,
diffusion-limited stage, where molecules of both species
diffuse from the bulk
with a $t^{1/2}$ time dependence (Fig.~4).
During this second stage (discussed above in sec. 4.3), 
the coverage is dominated by the more mobile
species having a faster diffusion.
In cases where the less mobile species is more surface active, a third stage
is predicted. Here one species undergoes desorption,
while the coverage gradually
becomes dominated by the second, energetically favorable
surfactant. This competition can be seen
in Fig.~1.
The final relaxation towards equilibrium is usually diffusion-limited,
exhibiting an asymptotic $t^{-1/2}$ behavior 
(Fig.~3), which was explained
in sec. 4.1.

For surfactant
mixtures exhibiting kinetically limited adsorption, we find
a ``synergistic'' effect, where the mixture kinetics may be  
considerably different from that of the individual species.
In cases of high equilibrium surface coverage, our model predicts 
a significant increase in
the limiting time scale due to coupling between the two
surfactants.

We have managed to relate the kinetic behavior of the two-surfactant
mixture with the properties of its individual constituents,  [eq.~(\ref{mix})].
The kinetic behavior of the mixture can be predicted based on equilibrium
isotherms and single-surfactant data.
Our results are in good agreement with experiments.
However, since previous experiments were restricted to surfactants of very
different adsorption time scales, our theory could be checked only
in those experimental conditions.
Further experiments on surfactant
mixtures, especially with comparable adsorption
time scales and equilibrium surface coverages, are needed
in order to get a better verification of our model.

The problem we have dealt with in this work is another example of the ease at
which the free-energy
approach to adsorption kinetics can be generalized.
Examples for further extensions are, for instance,
the addition of adsorption barriers and treatment of
surfactant adsorption from micellar
solutions.

\vspace{2cm}
\newlength{\tmp}
\setlength{\tmp}{\parindent}
\setlength{\parindent}{0pt}
{\em Acknowledgments}
\setlength{\parindent}{\tmp}

We thank E. Franses,
D. Langevin, R. Miller and C. Radke for
stimulating discussions and correspondence.
Partial support from the Israel Science Foundation founded by
the Israel Academy of Sciences and Humanities -- Centers of
Excellence Program -- and the U.S.-Israel Binational Foundation
(B.S.F.) under grant no.~94-00291 is gratefully acknowledged.

\pagebreak
\section*{Nomenclature}

Below is a summary of the symbols used in this work.
In order to make the formulation as concise as possible,
we have extensively used dimensionless forms.
Whenever necessary, a translation to the more practical,
dimensional quantities is provided, using the symbols of
{\it e.g.} ref.~\cite{FainermanMiller}.

\subsection*{Symbols}

\begin{tabular}{lcl}

$a$ &---& molecular dimension (equal to the inverse square root
of the maximum \\
& & surface density, $\Gamma_\infty^{-1/2}$)
 \\
$D$ &---& diffusion coefficient \\
$\Delta f$ &---& excess free energy per unit volume of
the bulk solution \\
$f_0$ &---& interfacial contribution to the free energy
per unit area \\
$t$ &---& time \\
$x$ &---& distance from the interface \\ 
$\alpha$ &---& energy (in units of $k_{\rm B}T$) gained by a 
surfactant molecule by migrating to \\
& & the interface \\
$\beta$ &---& energy (in units of $k_{\rm B}T$) of lateral 
attraction between two surfactant \\
& &  molecules of the same species \\
$\Delta\gamma$ &---& change in interfacial tension \\
$\varepsilon$ &---& energy (in units of $k_{\rm B}T$) of lateral 
attraction between two surfactant \\
& &  molecules of different species \\
$\mu$ &---& chemical potential (in units of $k_{\rm B}T$) \\
$\tau$ &---& time scale of a kinetic process \\
$\phi$, $\psi$ &---& volume fractions of the two surfactant
species (equal to the corresponding \\
& & concentrations multiplied by $a^3$; $\phi=a^3 c_{\rm A}$,
$\psi=a^3 c_{\rm B}$) \\
$\phi_0,\psi_0$ &---& surface coverages of the two surfactant
species (equal to the corresponding \\
& & surface densities multiplied by $a^2$; 
$\phi_0=a^2\Gamma_{\rm A}$, $\psi_0=a^2\Gamma_{\rm B}$) \\
$\bar{\phi}_0,\bar{\psi}_0$ &---& surface coverages in
the single-surfactant solutions

\end{tabular}

\subsection*{Subscripts}

\begin{tabular}{lcl}

A,B &---& corresponding to surfactant A, B \\
0 &---& value at the interface \\
1 &---& value at the sub-surface layer of the solution \\
b &---& value at the bulk reservoir \\
eq &---& value at equilibrium \\ 

\end{tabular}


\pagebreak
\section*{Appendix: Numerical Scheme }
\label{scheme}
\setcounter{equation}{0}

The objective of the numerical scheme is to
solve the set of four equations --- the two Ward-Tordai equations (\ref{a11})
and the two interfacial ones (\ref{a12}).
The hardcore of the problem is the evaluation of the integral in eq.~(\ref{a11}),
\[
   \ap_{0} (t) = \frac{1}{ a } \sqrt{\frac{ \DA }{\pi}}
\left[ 2\phi_{\rm b} \sqrt{t} - \int_{0}^{t} \frac{ \phi_{1}
(\tau) }{ \sqrt{t-\tau} }{\rm d}\tau  \right] + 2\phi_{\rm b} -
\phi_{1} 
\]
This is done by discretizing the time variable, 
$ t  $, and transforming the integral into a
recursion relation \cite{numeric}.
The singularity in the integrand is removed by substituting
$ x = \sqrt{ t - \tau } $.
Evaluating the integral at discrete time steps, 
$ \Delta t = t_{ k+1 } - t_k $,
we obtain
\[
   \int_{0}^{t} \frac{ \phi_1
(\tau) }{ \sqrt{t-\tau} } {\rm d} \tau \rightarrow \sum_{k=1}^{n}
\left[ \phi_1 ( t_n - t_k ) + \phi_1 (t_n - t_{k-1} ) \right]
( \sqrt{ t_{k} } - \sqrt{ t_{k-1} } )
 \]
where $ t_k \equiv k \Delta t $ and $ t_n \equiv t $. 
The initial conditions are
$\ap_0(0)=\ap_1(0)=\ap_b \;$ and $\bp_0(0)=\bp_1(0)=\bp_b $.
Adding the rest of the terms in the Ward-Tordai equations, 
we obtain linear relations
between $\phi_0$ and $\phi_1$, and between $\bp_0$ and $\bp_1$.
Substituting these relations into the kinetic boundary 
conditions (\ref{a12})
and using a
finite difference to evaluate the first time derivative,
$ \partial \ap_0 / \partial t \rightarrow [ \ap_0 ( t_{k+1} ) -
\ap_0 ( t_k ) ] / \Delta t $,
a set of four algebraic equations is obtained, relating
the values of $\ap_0$, $\bp_0$, $\ap_1$ and $\bp_1$
at the current time step, $ t = t_n $,
\begin{eqnarray*}
\ap_0 ( t_n ) & = &
\frac{ 1}{ a } \sqrt { \frac { \DA  }{ \pi } } \left[ 2 
\ap_{\rm b} \sqrt {t_n} -
 \sum_{k=1}^{n-1}
 \phi_1 ( t_n - t_k )
( \sqrt{ t_{k+1} } - \sqrt{ t_{k-1} } ) - \ap_0 ( \sqrt{t_n} - 
 \sqrt{ t_{n-1}} )\right]  \\
& & + \, 2 \ap_{\rm b} - ( 1 + \frac{ 1}{ a } \sqrt 
{ \frac { \DA  }{ \pi } } 
 \sqrt{\Delta t} ) \ap_1(t_n)
\\ \nonumber \\
 \ap_0 ( t_n ) & = & \ap_0 ( t_{n-1} ) +
\frac{ \DA }{ a^2 } \Delta t  \left\{
\ap_1 ( t_n ) \ln \left[
 \frac{ \water_0 ( t_n ) \ap_1 ( t_n ) }{ \ap_0 ( t_n ) } 
\right] +
\ala + \bta \ap_0 ( t_n )+ \varepsilon \bp_0 ( t_n ) \right\}
\label{boundryCond}
\end{eqnarray*}
and similar two equations for surfactant B.
This set of equations is solved numerically, using a
binary search method.


\pagebreak
\setlength {\baselineskip} {12pt}


\vspace{1cm}
\setlength {\baselineskip} {16pt}

\section*{Table Caption}

\begin{itemize}

\item[{\bf Table 1}]
Comparison of the predicted
$\sqrt{\tau_\gamma}$ [eq.~(\ref{mix})]
with experimental results \cite{FainermanMiller}.
The materials used were sequences of Triton X
mixtures.  The single-surfactant values, 
$\bar{\ap}_0$, $\bar{\bp}_0$, 
$ \Delta\bar{\gamma}_{\rm A}\sqrt{\bar{\tau}_{\rm A}} $,
$ \Delta\bar{\gamma}_{\rm B}\sqrt{\bar{\tau}_{\rm B}} $,
and equilibrium isotherms for the mixture,
$\ap_{\eq}$ and $\bp_{\eq}$, were taken from the same reference.
The values for $\Delta\gamma\sqrt{\tau_\gamma}$ 
(given in units of dyne~sec$^{1/2}$/cm)
are obtained experimentally
from the asymptotic slope of the $\gamma$ vs. $t^{-1/2}$ curves [see
eq.~(\ref{AsyForm})].
The predicted values for $\Delta\gamma\sqrt{\tau_{\gamma}}$ 
of the mixture and
the corresponding experimental results are given in the
columns indicated by `th' and `exp', respectively. 
The last column shows
the respective error between theory and experiment.

\end{itemize}

\section*{Figure Captions}

\begin{itemize}

\item[{Fig. 1}]
Surface coverage in a mixture of interacting surfactants.
The dotted, dashed and solid lines are the surface coverages
of surfactants {\rm A} ($\ap_0$), {\rm B} ($\bp_0$) and the total
coverage ($\ap_0+\bp_0$),
respectively. The assigned parameters are:
$\ap_{\rm b} = 1\times 10^{-4} $, $ \bp_{\rm b} = 2\times 10^{-4} $,
$\ala = 10 $, $\alb = 9 $,
$\bta = \btb = 3 $,
$\varepsilon = 1$,
$ \sqrt{ \DA } / a = 300 \; {\rm sec}^{-1/2} $, 
$ \sqrt{ \DB } / a = 900 \;
{\rm sec}^{-1/2} $.
This implies that surfactant {\rm A} diffuses more slowly but
is more surface active ($\ala > \alb$).

\item[{Fig. 2}]
Dynamic surface tension for the system of fig.~1.

\item[{Fig. 3}]
Dynamic surface tension for the system of fig.~1, redrawn for long
times. The curve exhibits the $t^{-1/2}$ asymptotic behavior characteristic
of DLA.

\item[{Fig. 4}]
Surface coverage (a) and dynamic surface tension (b) 
for the system of fig.~1, redrawn for
intermediate short times. 
The curve exhibits the expected $t^{1/2}$ behavior characteristic
of DLA at short times.

\item[{Fig. 5}]
Dynamic surface tension for the system of fig.~1, redrawn for
extremely short times. The curve exhibits the predicted linear
behavior.

\item[{Fig. 6}]
Comparison between the numerical solution of the complete set of
equations (solid line) and the analytic solution assuming a
linear adsorption isotherm (dashed line).
The parameters used in the calculations are: $\ap_{\rm b}=2 \times 10^{-5}
$, $ \bp_{\rm b} = 10^{-4}$, $ \ala = 6 $, $ \alb = 5 $,
$ \bta = \btb = \varepsilon = 0
$, $ \sqrt{\DA}/a = 2 \times 10^4 \; {\rm sec}^{-1/2} $ and
$ \sqrt{\DB}/a = 10^4 \; {\rm sec}^{-1/2} $.
The linear
approximation is valid only in the beginning of the process, and
fails when the coverage becomes large.

\end{itemize}


\vspace{3cm}
\begin{table}[tbh]
\begin{center}
\begin{tabular}{|c|c||c|c|c|c|c|c|c|}
\hline
&&&&&&&& \\
  $ {\rm A} $
& $ {\rm B} $
& $ \ap_{\seq} / \bar{\ap}_{\seq} $
& $ \bp_{\seq} / \bar{\bp}_{\seq} $
& $ \Delta\bar{\gamma}_{\rm A}\sqrt{ \bar{\tau}_{\rm A} } $
& $\Delta \bar{\gamma}_{\rm B}\sqrt{ \bar{\tau}_{\rm B} } $
& $ \Delta\gamma\sqrt{\tau_\gamma} $~(th)
& $ \Delta\gamma\sqrt{\tau_\gamma} $~(exp) &
error
\\
&&&&&&&& \\
\hline
\hline
&&&&&&&& \\
X-405 & X-45 & 0.13 & 0.69& 0.6 & 62 & 29.5 & 32 & 8\% \\
&&&&&&&& \\
X-405 & X-100 & 0.25 & 0.67 & 0.6 & 38 & 17.1 & 17 & 0.6\% \\
&&&&&&&& \\
X-405 & X-114 & 0.06 & 0.71& 0.6 & 14 & 7.1 & 6.8 & 4\% \\
&&&&&&&& \\
X-405 & X-165 & 0 & 1.4 & 0.6 & 4.4 & 8.6 &  6.5 & 33\% \\
\hline
\end{tabular}
\end{center}
\caption[]{}
\end{table}


\pagebreak

\begin{figure}[p]
\epsfysize=16\baselineskip
\centerline{\hbox{ \epsffile{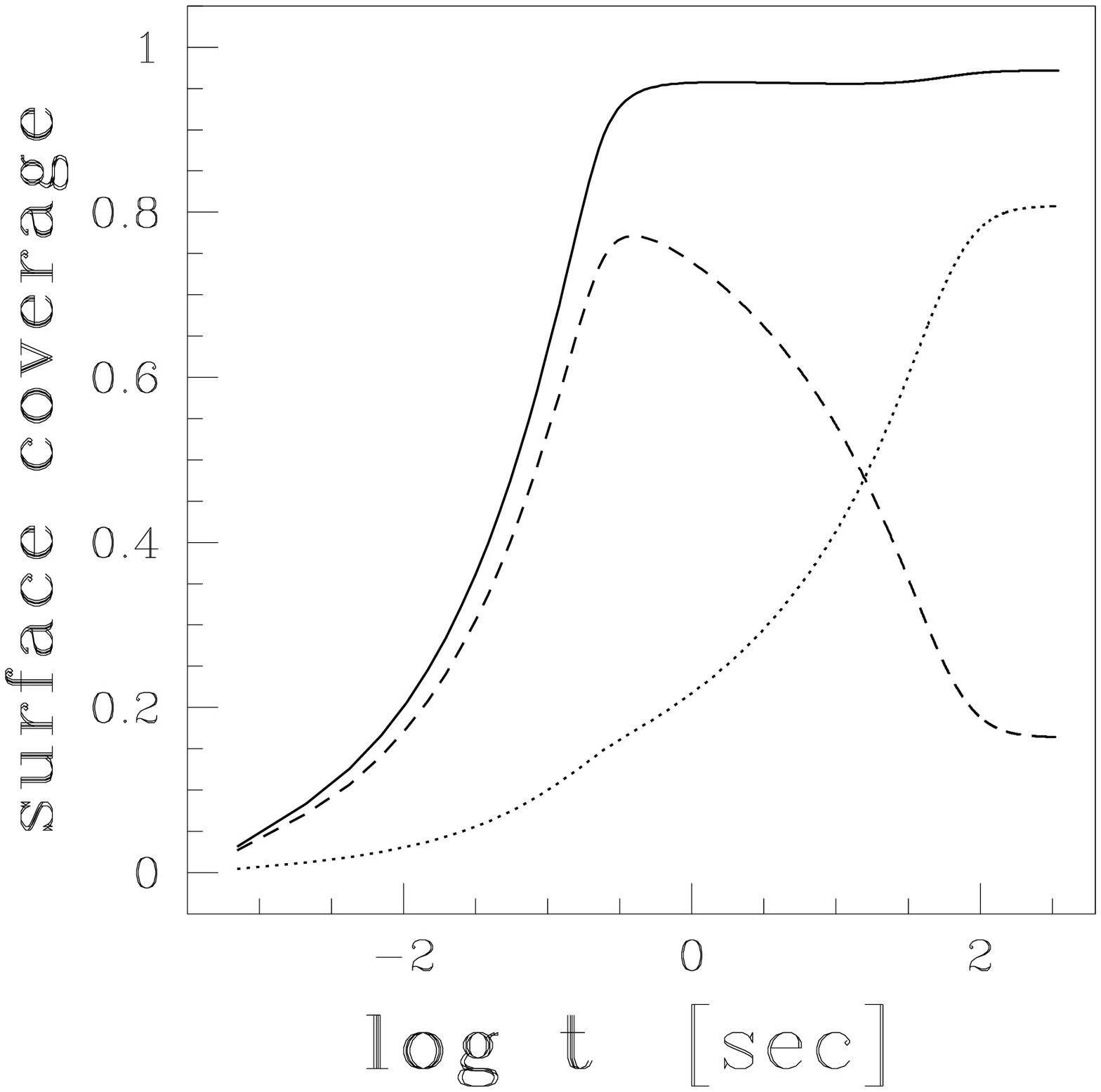} }}
\caption[]{}
\label{fig1}
\end{figure}

\begin{figure}[p]
\epsfysize=16\baselineskip
\centerline{\hbox{ \epsffile{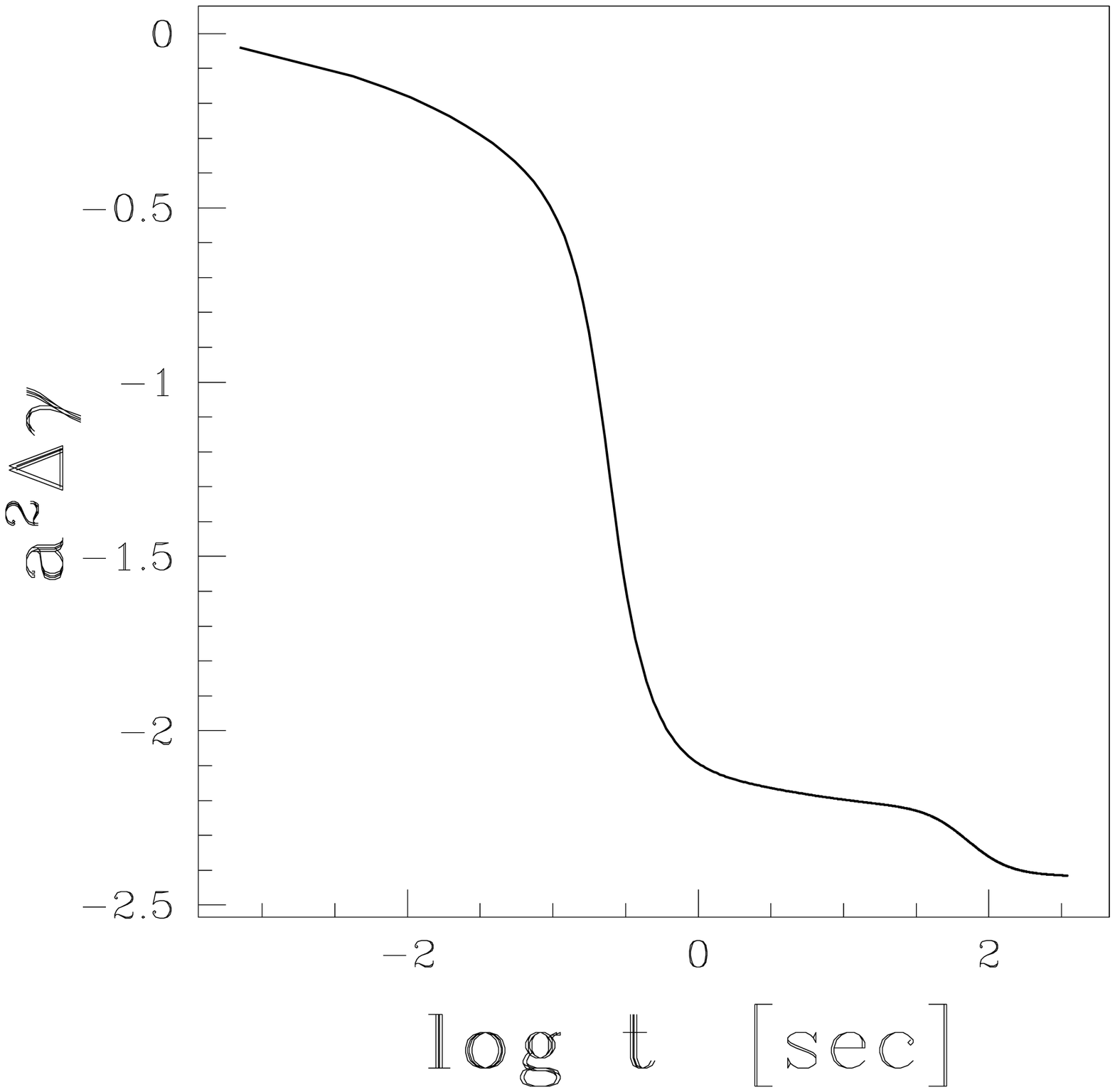} }}
\caption[]{}
\label{fig2}
\end{figure}

\begin{figure}[p]
\epsfysize=16\baselineskip
\centerline{\hbox{ \epsffile{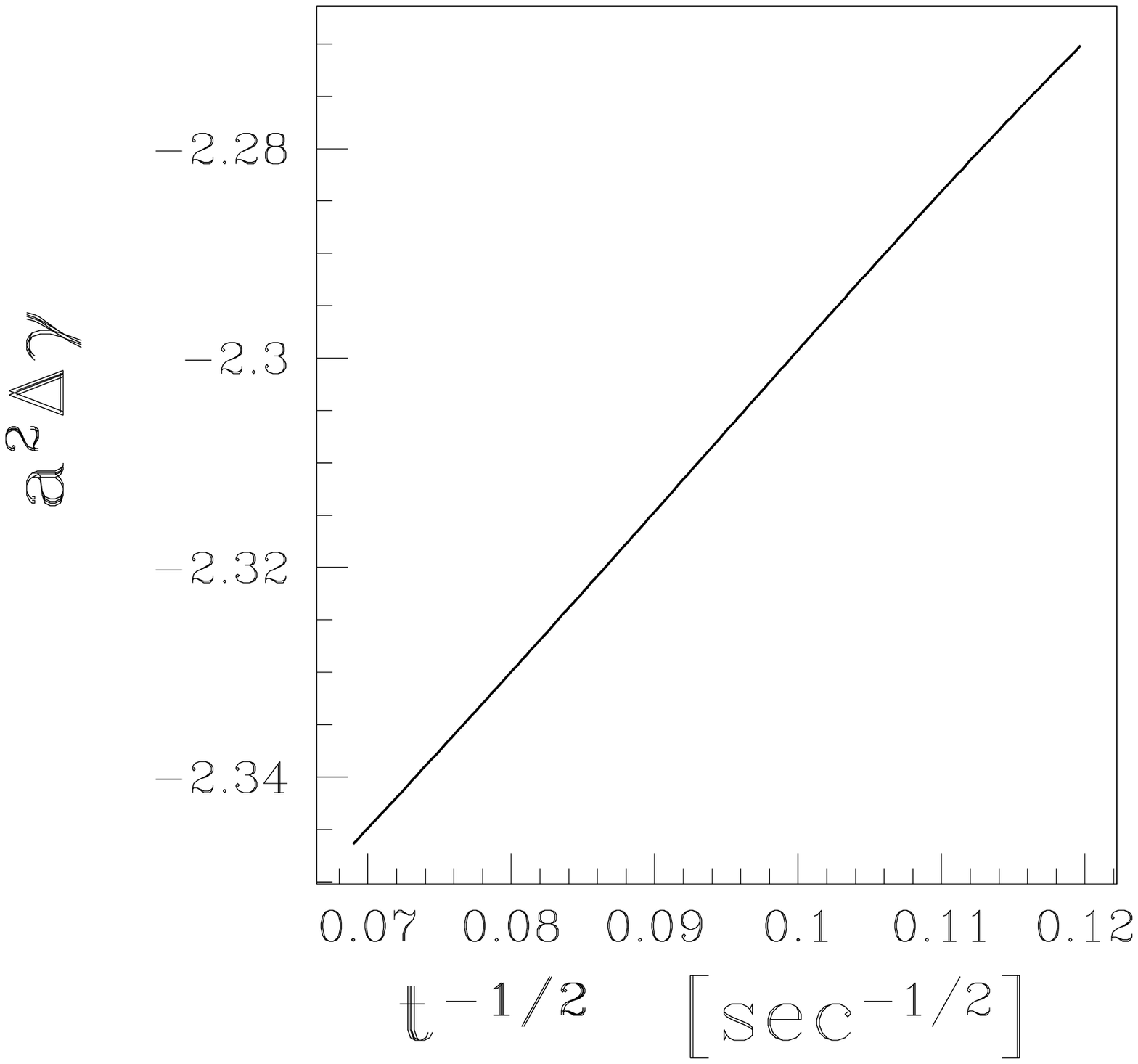} }}
\caption[]{}
\label{fig3}
\end{figure}

\begin{figure}[p]
\epsfysize=16\baselineskip
\centerline{\hbox{\epsffile{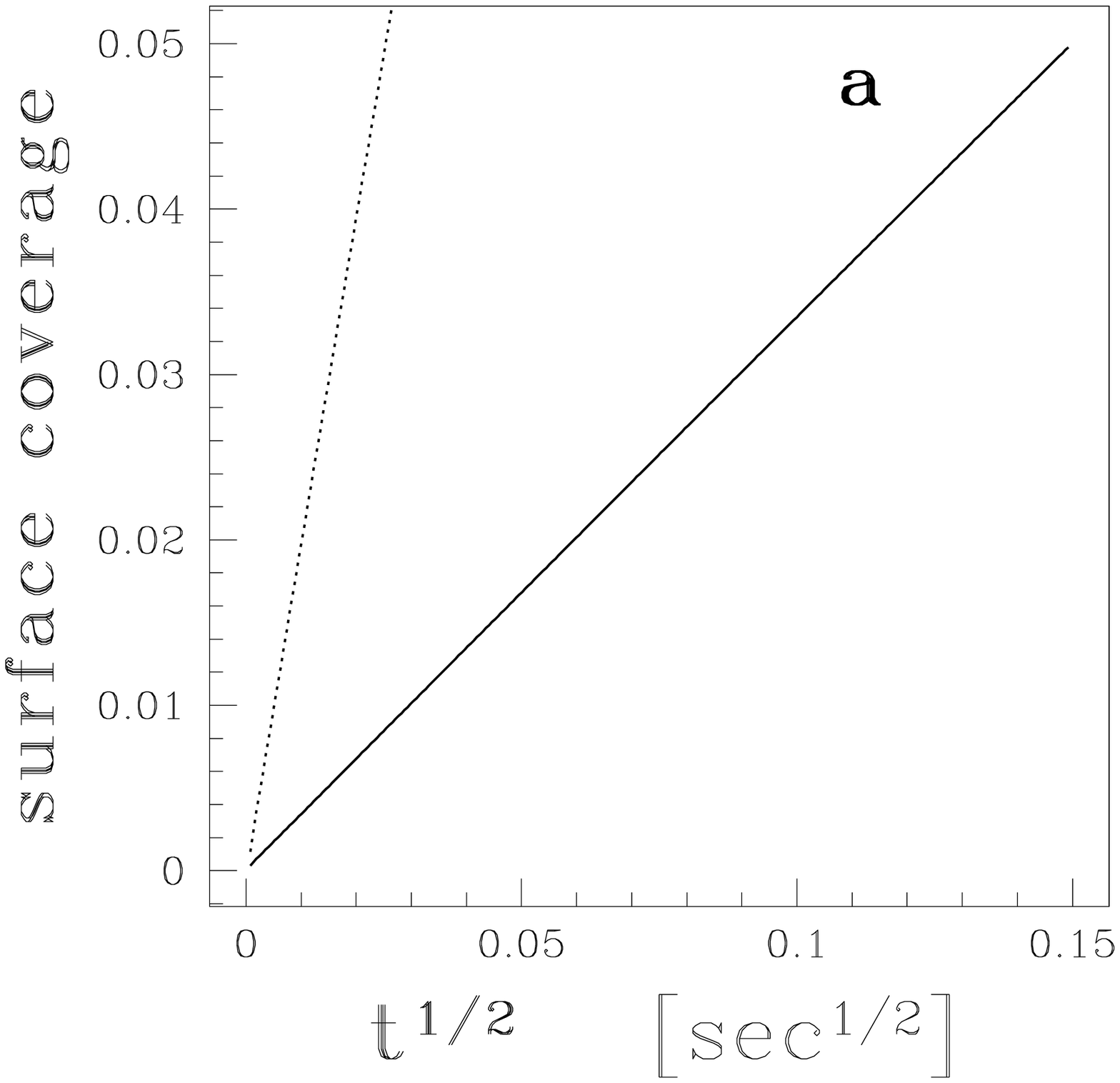}}
\epsfysize=16\baselineskip \hbox{\epsffile{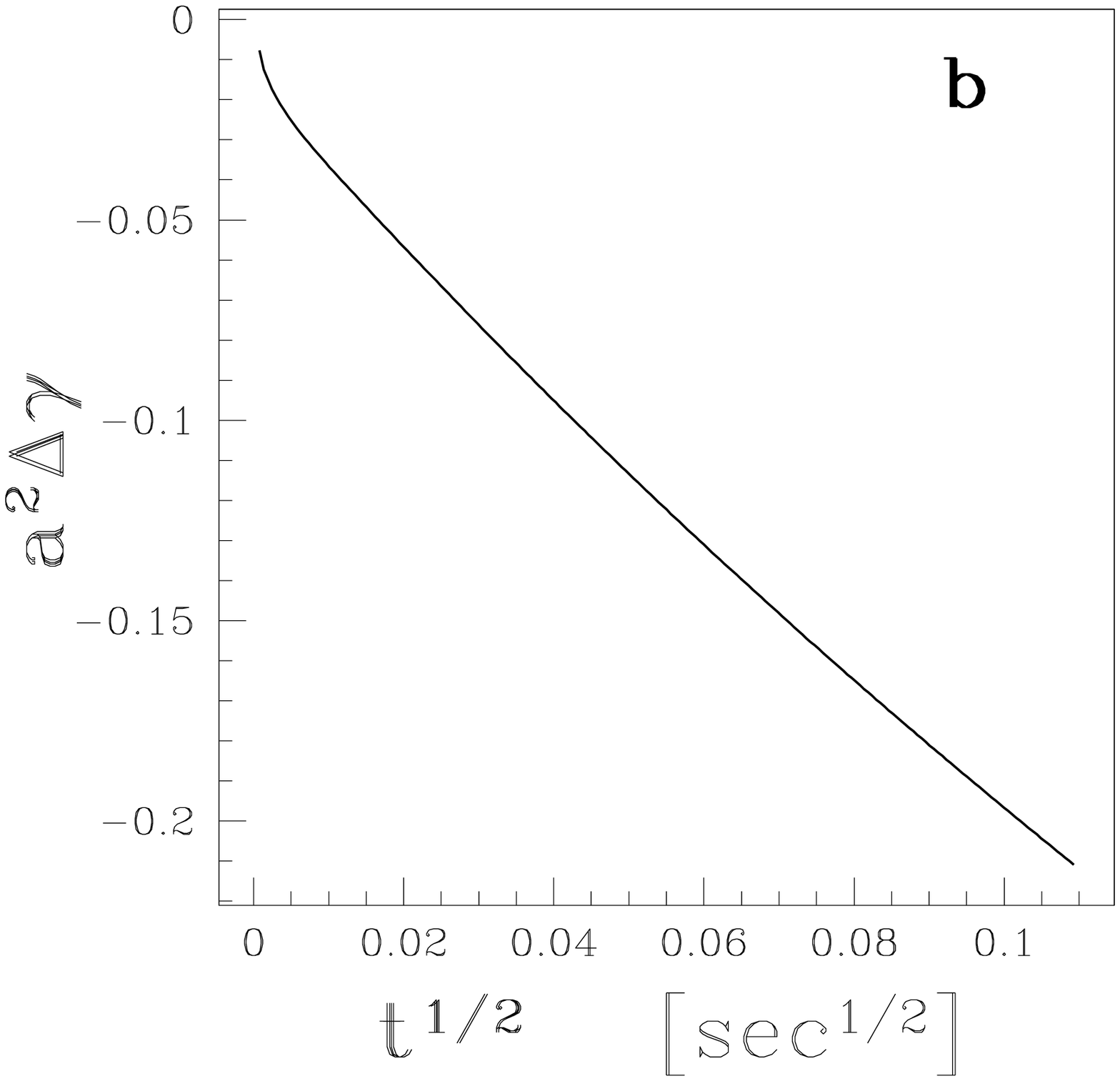}}}
\caption[]{}
\label{fig4}
\end{figure}

\begin{figure}[p]
\epsfysize=16\baselineskip
\centerline{\hbox{ \epsffile{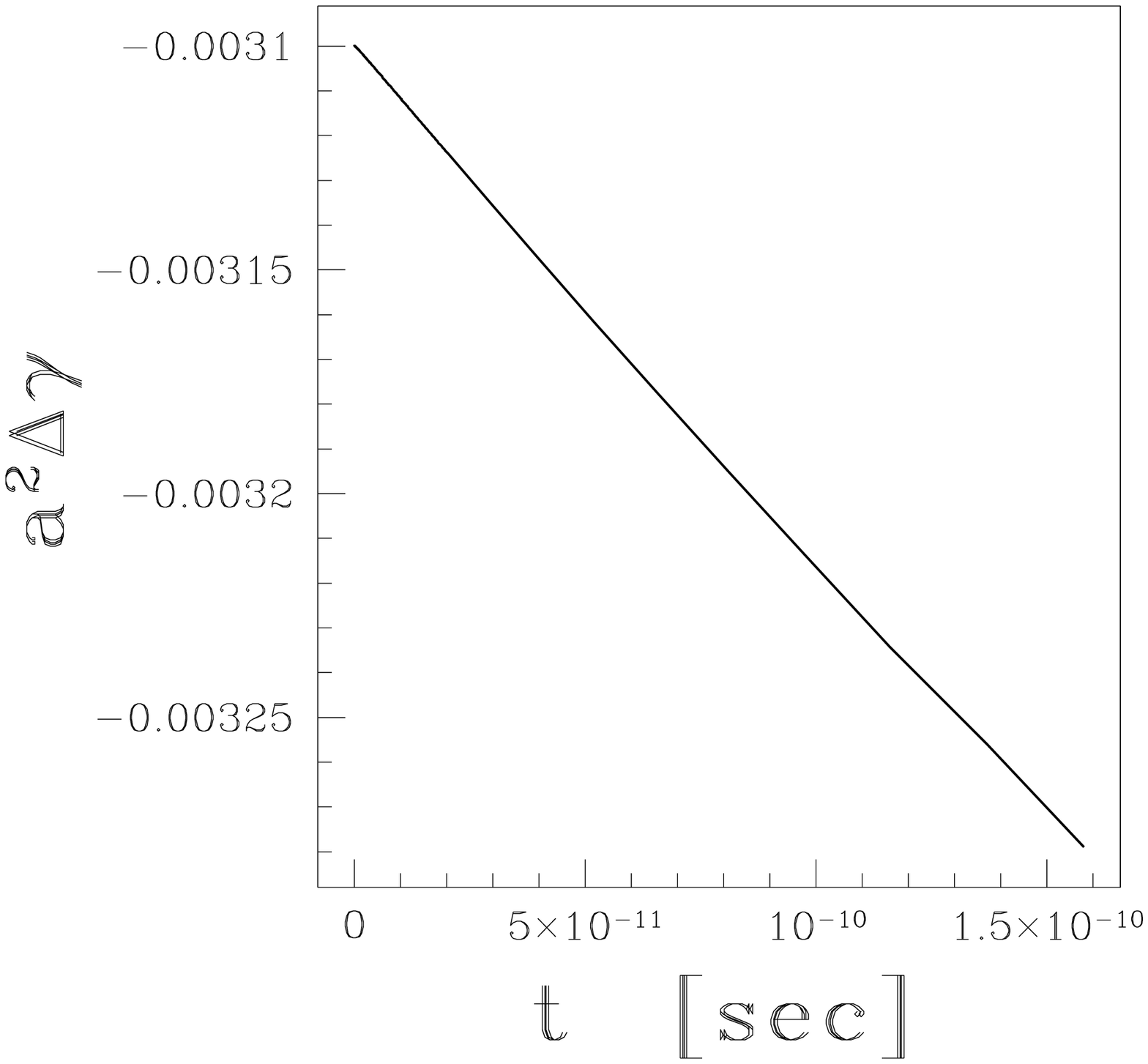} }}
\caption[]{}
\label{fig5}
\end{figure}

\begin{figure}[p]
\epsfysize=16\baselineskip
\centerline{\hbox{ \epsffile{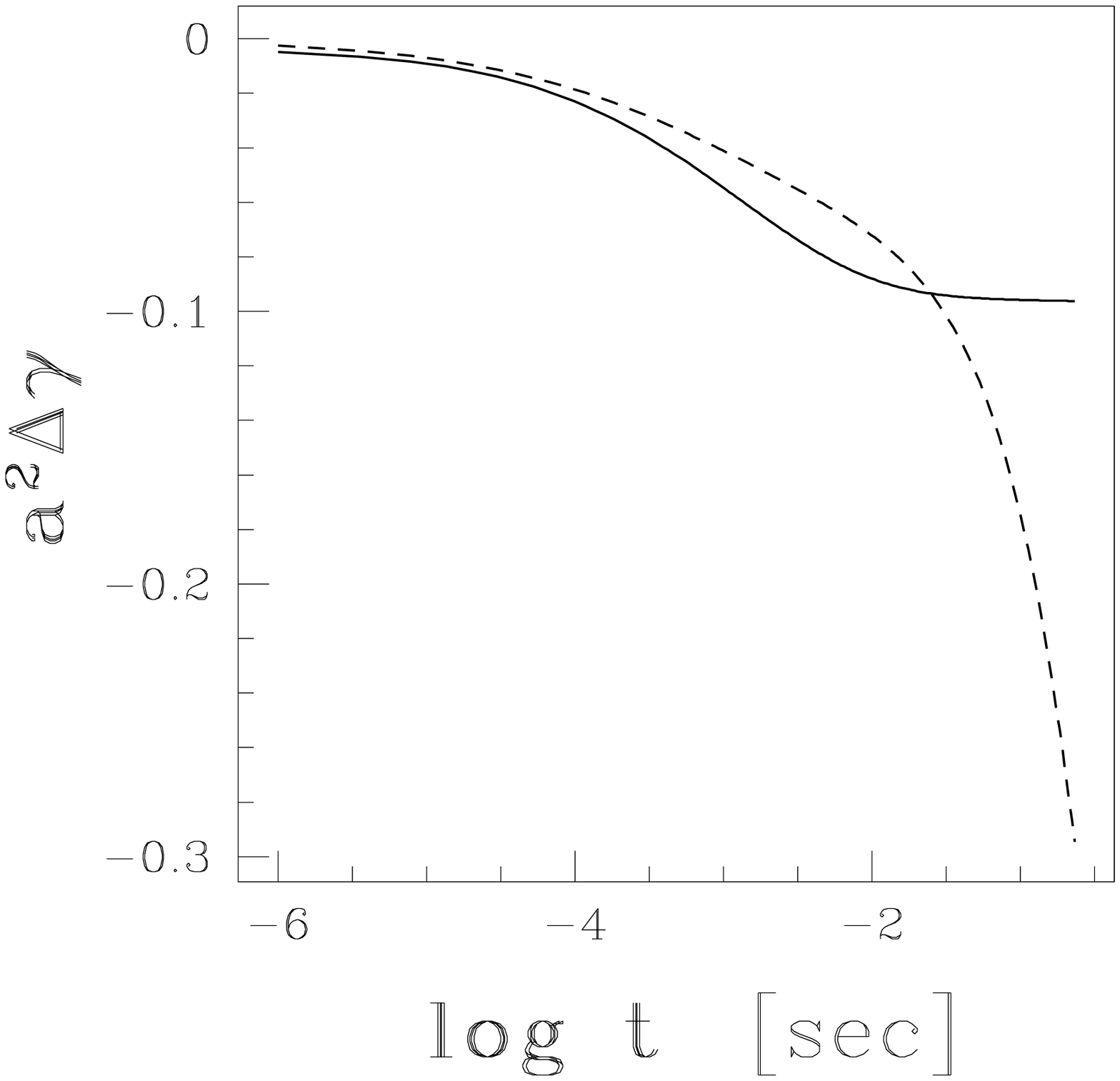} }}
\caption[]{}
\label{fig6}
\end{figure}



\begin{thebibliography}{99}
\bibitem{micelle} It should be noted, however, that industrial 
applications often involve surfactant solutions close to and above
the cmc. For equilibrium theories of such surfactant mixtures, 
see refs.~\cite{Clint,Blank2}.
\bibitem{GammaExp} Note that this is a {\rm tertiary} system
   (two solutes in a solvent), requiring three parameters for a 
complete
   description of the interactions (in our case $\bta$, $\btb$
   and $\varepsilon$).
\end{thebibliography}
\end{document}